# Cooperative Strategies for Wireless-Powered Communications: An Overview

He Chen[1], Chao Zhai[2], Yonghui Li[1], and Branka Vucetic[1]

[1] School of Electrical and Information Engineering, The University of Sydney, Australia
Email: {he.chen, yonghui.li, branka.vucetic}@sydney.edu.au
[2] School of Information Science and Engineering, Shandong University, China
Email: chaozhai@sdu.edu.cn

*Abstract*-Radio frequency (RF) energy transfer and harvesting has been intensively studied recently as a promising approach to significantly extend the lifetime of energy-constrained wireless networks. This technique has a great potential to provide relatively stable and continuous RF energy to devices wirelessly; it thus opened a new research paradigm, termed wireless-powered communication (WPC), which has raised many new research opportunities with wide applications. Among these, the design and analysis of cooperative schemes towards efficient WPC have attracted tremendous research interests nowadays. This article provides an overview of various cooperative strategies for WPC, with particular emphasis on relaying protocols for wireless-powered cooperative communications, cooperative spectrum sharing schemes for cognitive wireless-powered networks, and cooperative jamming strategies towards wireless-powered secure communications. We also identify several interesting research directions in this area before concluding this article.

## I. INTRODUCTION

Radio frequency (RF) energy transfer and harvesting technique has recently emerged as a promising way to extend the lifetime of energy-constrained wireless networks, especially when the conventional energy harvesting techniques from renewable energy sources are not applicable. This technique enables wireless terminals to scavenge energy from RF signals broadcast by ambient/dedicated wireless transmitters to support their operation and information transmission. This new communication format has been termed as wireless-powered communication (WPC) in the literature (see [1] and references therein), which advocates the dual function of RF signals for both information delivery and energy transfer.

In WPC, wireless terminals can avoid being interrupted by their batteries' depletion, which can thus be deployed more flexibly and maintained in lower costs. In this sense, WPC has great potentials to sustain the network operation than its conventional battery-powered counterpart in a long run. Thanks to these inherent merits, WPC has been regarded as an indispensable and irreplaceable building block of a wide range of applications (e.g., RFID, wireless sensor networks, machine-to-machine communications, low-power wide-area networks, and Internet of Things, etc.).

On the other hand, as an effective means to combat the severe propagation loss of wireless links, cooperative communication techniques have drawn tremendous research interests in the past decades. Their basic idea is to allow single-antenna devices to share their antennas and work collaboratively such that a virtual MIMO system can be constructed to realize the space diversity. Consequently, the overall communication quality, including reception reliability, energy efficiency, system capacity and network coverage, can be dramatically improved.

Cooperative techniques have also been integrated into cognitive radio networks and secure communications. Various cooperative spectrum sharing protocols have been developed to stimulate the cooperation between primary users (PUs) and secondary users (SUs), where the SUs relay traffics for the PUs, in exchange for the access of licensed spectrum to fulfill their own transmission requirements. Besides, to enhance the physical layer security of wireless networks, various cooperative jamming schemes have been designed to eliminate the information leakage of legitimate users in the presence of malicious eavesdropper(s).

A fundamental hurdle for the wide proliferation of WPC is the high attenuation of RF signals over distance. As such, the amount of harvested energy at wireless-powered devices is normally very limited, which significantly confines the performance of WPC. Multiple-antenna techniques can be implemented to overcome this hurdle. However, this solution may be infeasible due to the size and cost limitations of wireless-powered devices in many applications. In this context, cooperative communication techniques stand out as a very attractive solution to boost the efficiency of WPC since it can enable multiple single-antenna nodes to work collaboratively to realize spatial diversity.

In WPC, RF signals are employed to deliver both information and energy. As such, besides the conventional cooperation for information transmission, the cooperation for energy transfer should also be implemented to overcome the severe propagation attenuation caused by path-loss and channel fading. In this sense, *how and when to cooperate in WPC becomes a more complex question to deal with*. Furthermore, since the harvested energy is normally limited, when to utilize it to perform an appropriate transmission/cooperation becomes critical to the system performance. As such, the existing cooperative strategies may no longer be efficient for WPC. These strategies should be revisited and some totally new cooperative strategies should be developed for efficient WPC.

The main objective of this article is to provide a holistic overview of the state-of-the-art of cooperative strategies towards efficient WPC, with particular interests in cooperative relaying, cooperative spectrum sharing, and cooperative jamming techniques for wireless-powered relay, cognitive radio, and secure communication networks. Furthermore, we will identify several promising yet challenging research trends in these areas.

## II. Relaying Protocols for Wireless-powered Cooperative communications

A classical three-node cooperative communication network (CCN) consists of one source S, one relay R and one destination D. S intends to transmit its information to D with the assistance of R, which processes the signal received from S and forwards it to D via various relaying protocols, such as amplify-and-forward (AF) and decode-and-forward (DF), etc. In this article, we refer to a CCN with one or more wireless-powered nodes as a wireless-powered CCN (WPCCN). Considering whether only S or R or both nodes are powered by the RF energy[1], WPCCNs have several elementary network configurations. We subsequently describe each configuration and several representative work:

### A. WPCCN with Wireless-powered Source Only

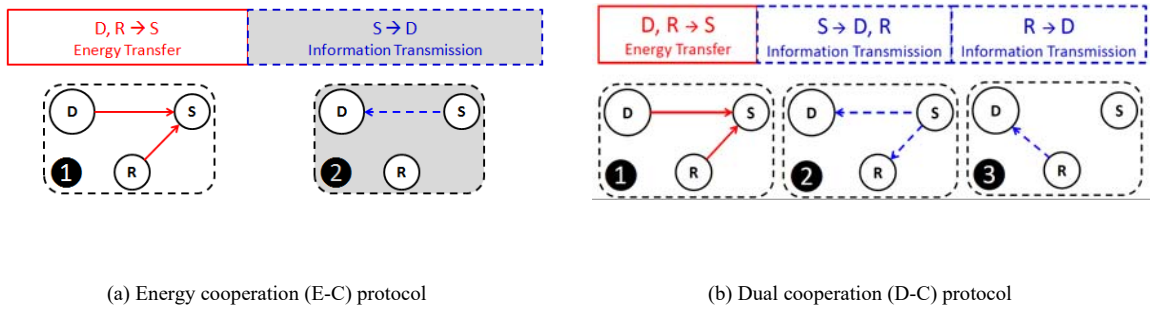

(a) Energy cooperation (E-C) protocol      (b) Dual cooperation (D-C) protocol

Fig. 1 Block diagrams of energy cooperation (E-C) and dual cooperation (D-C) protocols [2].

In this setup, only S has no embedded energy supply and is purely powered by the RF energy, while both R and D have stable and continuous power supply, e.g., they may operate with large-capacity batteries or connect to the power grid. As such, S needs to harvest RF energy from D before transmitting its information to D. In this case, R can not only assist S to forward its information to D as in conventional CCNs, but also can help D charge S during the wireless energy transfer phase. Reference [2] investigated this type of WPCCNs and developed two cooperative protocols, namely energy cooperation (E-C) and dual cooperation (D-C), which differs in the relay's roles during each transmission block. The block diagrams of these two protocols are depicted in Fig. 1, in which we can see that both E-C and D-C split each transmission block into two phases for energy transfer and information transmission, respectively. In the D-C protocol, the information transmission phase is further divided into two consecutive time slots to enable the information relaying.

---

[1] Note that the destination D is normally not considered as a wireless-powered node as it does not perform any information transmission.

As shown in [2], the E-C and D-C protocols can be superior to each other depending on network setups. More specifically, the E-C protocol achieves higher throughput at high signal-to-noise ratio (SNR), especially when S and R are located far from each other. On the other hand, when the SNR is not high, and S and R are close to each other, the D-C protocol outperforms the E-C protocol in terms of system throughput. In this sense, an adaptive protocol that switches between E-C and D-C protocols adaptively can further enhance the system performance.

It is worth noting that the E-C protocol can be designed and implemented more easily compared to the D-C protocol. Specifically, R only needs to act as a wireless charger in the E-C protocol, while it is required to perform both wireless energy transfer and information relaying in the D-C protocol. Furthermore, the optimal design of the E-C protocol involves the optimization of network parameters including the time allocation factors of the energy transfer phase and information transmission phase, the transmit powers of R and D. In the D-C protocol, in addition to those network parameters in the E-C protocol, the transmit power of R in the information forwarding phase should be optimized jointly with the transmit power of R in the energy transfer phase and the time allocation factors.

These two cooperative schemes can find their potential applications in machine-to-machine (M2M) communications with data aggregation. It is estimated that the number of M2M connections may reach 200 billion by 2020. Due to the massive number of machine-type devices (MTDs) to be deployed and their inherent low traffic, it becomes appealing to make them wireless-powered such that the huge cost of frequent battery replacement/recharging can be saved and their continuous operations can be sustained. Furthermore, concurrent transmissions from enormous MTDs to the core network (e.g., the base station (BS) in cellular networks) can lead to severe radio access network congestion. One promising solution to tackle this massive connectivity issue is to implement the data aggregation technique. Specifically, data aggregators are deployed across the whole network such that the MTDs first send their traffic to a designated data aggregator, which then forwards the collected packets to the BS. In this application, S is a MTD, R is a data aggregator and D is a BS; the data aggregator and BS work collaboratively to support the uplink transmission of the wireless-powered MTD.

*B.  WPCCN with Wireless-powered Relay Only*

As for this network configuration, S has constant power supply, while R is a wireless-powered node or it has embedded energy source but is not willing to use its own energy to help S. In this context, S should transfer a certain amount of energy wirelessly to the relay before asking the relay to forward its data to D. This actually provides an effective mechanism to motivate R to help S. As a seminal work in this area, reference [3] developed two new relaying protocols based on the receiver structures adopted at R, termed time switching-based relaying (TSR) and power splitting-based relaying (PSR) shown in Fig. 2. In TSR, the energy transfer and information transmission from S to R occurs in different time slots and R only needs to switch between the units of EH and information decoding (ID). In contrast, the energy transfer and information transmission from S to R are performed concurrently in PSR. To this end, R needs a power splitting component to split the received signal in the power domain: one part for EH and the other part for ID. The simulation results in [3] revealed that PSR always outperforms TSR in terms of system throughput. This is understandable since PSR has a higher spectral efficiency. However, compared to PSR, TSR has a lower hardware complexity and cost. In terms of their optimal system designs, the TSR protocol requires to optimize the time switching factor and the power allocation between energy transfer and information transmission for S. When it comes to the PSR protocol, only the power splitting factor needs to be optimized since the optimal transmit power for S is apparently its maximum allowed power.

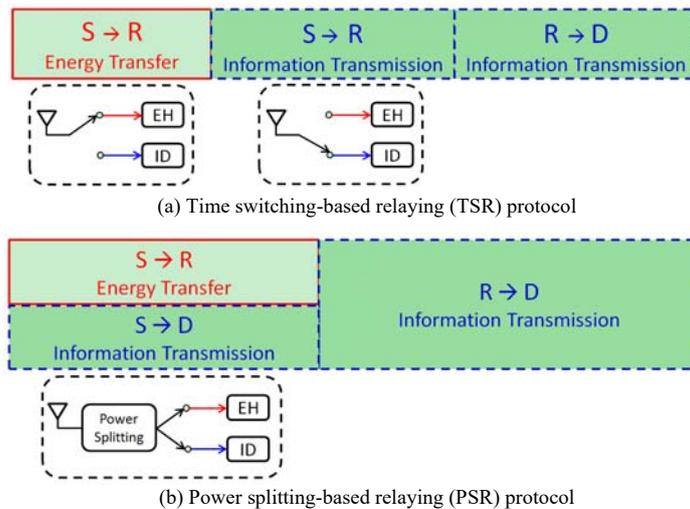

(a) Time switching-based relaying (TSR) protocol

(b) Power splitting-based relaying (PSR) protocol

Fig. 2 Block diagrams of TSR and PSR protocols and their corresponding receiver structures at the R [3].

The practical application of the TSR and PSR protocols implicitly requires that S has plenty of energy to perform both effective energy transfer and information transmission. As such, the downlink transmission from a (small cell) BS, which is connected to the power grid, to a cell-edge device via an intermediate node can be a typical application. The intermediate node can be a wireless-powered node or has embedded energy supply but is unwilling to consume its own energy to help the BS. In both cases, the BS needs to first charge the intermediate node wirelessly before asking the node to forward its information to the cell-edge user.

*C. WPCCN with Wireless-powered Source and Relay*

When both S and R are wireless-powered nodes, they need to first harvest energy from D or another dedicated RF energy transmitter before transmitting/forwarding data. This type of WPCCN was studied in [4], wherein a harvest-then-cooperate (HTC) protocol was proposed and analyzed. In the HTC protocol, S and R first harvest energy from D in the downlink (DL) and then cooperatively transmit data in the uplink (UL), as shown in Fig. 3. The performance evaluation of the HTC protocol was also extended to a more general scenario with multiple wireless-powered relays and popular single relay selection schemes. It was illustrated in [4] that the proposed HTC protocol is substantially superior to the existing non-cooperative schemes (e.g., harvest-then-transmit (HTT) protocol) in terms of average system throughput, which improves further with more relays deployed between S and R. It should be noted that in general, R should be closer to D than S to ensure that the cooperation between S and R can bring performance gains. However, even in this case, the HTC protocol can be inferior to the HTT protocol, which occurs when the instantaneous channel gains of the link S-R or R-D is weak. In this sense, designing an adaptive protocol with an adequate criterion to switch between the HTC and HTT protocols can further improve the system performance.

The optimal design of the HTC protocol is relatively simple. Only the time allocation between the energy transfer and information transmission phases is to be optimized since it can be verified that all nodes should transmit with their respective maximum power to maximize the system throughput. We envision that the HTC protocol can find its potential realizations in wireless sensing and monitoring scenarios, one of the fundamental applications of the upcoming internet of things (IoT). For example, S and R can be two wireless-powered sensor nodes, while D is an information collector that also acts as a wireless energy transmitter. The information collector D wants to collect the sensed data from the sensor node S under the help of another surrounding node R, which has no scheduled data to report in the current transmission cycle.

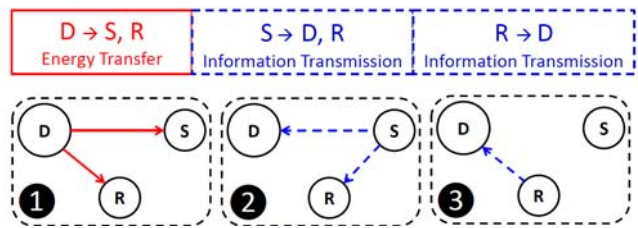

Fig. 3 Block diagram of the harvest-then-cooperate (HTC) protocol [4]

Finally, it is worth pointing out that the performance of wireless-powered relaying is nevertheless worse than that of the conventional relaying since the wireless power attenuates quickly during propagation. On the other hand, the intermediate relay nodes will be disconnected from the network if their batteries are depleted in conventional systems. In contrast, these relay nodes can still work under the support of harvested energy in WPC. In this sense, wireless-powered relaying can make the system operation more sustainable than its conventional counterpart.

## III. Cooperative Spectrum Sharing for Wireless-powered Cognitive Radio Networks

To meet the explosively-growing wireless data requirements, cognitive radio techniques have been intensively researched to enhance the utilization efficiency of licensed spectrum, by allowing SUs to access the spectrum belonging to PUs in an opportunistic manner. Spectrum sharing in cognitive radio networks (CRNs) can be generally clarified as three types: interweave, underlay, and overlay. Thanks to the space diversity brought by cooperative relaying, the overlay spectrum sharing, well-known as cooperative spectrum sharing (CSS), can achieve higher spectral efficiency than the non-cooperative interweave and underlay strategies by introducing certain coordination between PUs and SUs. The CSS is a win-win strategy for both primary and secondary systems as they actively seek chances to cooperate with each other to efficiently use the spectrum. More specifically, the SUs can act as cooperative relays to facilitate the primary data transmission with less resource consumption, as a return, they can access the licensed spectrum for their own data transmission in time, frequency, or space domain.

Incorporating WPC in the future CRNs introduced many new research opportunities. Particularly, the existing CSS strategies for conventional CRNs may no longer work properly in wireless-powered CRNs (WPCRNs) and thus should be re-designed, especially considering the randomness of EH amount, the efficiency of energy transfer, and the energy causality constraint, etc. In a WPCRN, both SUs and PUs could be wireless-powered. In the subsequent two subsections, we will discuss WPCRNs with wireless-powered SUs and PUs, respectively. In reality, the wireless-powered SUs and PUs can be IoT sensors without embedded energy supply, which may have no or have access to the licensed spectrum.

A. *WPCRN with Wireless-powered SU*

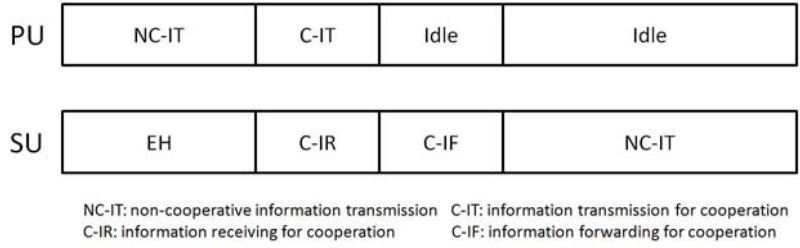

Fig. 4 Block diagram of a CSS protocol for WPCRNs with wireless-powered SU proposed in [5].

When a SU in a WPCRN has no fixed power supply (i.e., wireless-powered), it requires to first perform EH before serving as a relay for the primary system. During this EH period, the PU should rely on itself to transmit information to its receiver. Reference [5] proposed a novel time switching-based CSS strategy and maximized the SU's achievable throughput subject to the throughput constraint of the PU. In the proposed CSS protocol of [5], each transmission block is divided into four time slots with specific purposes, as illustrated in Fig. 4. The SU splits the harvested energy into two portions: one is used to cooperatively relay the PU's data to exchange for the opportunity of spectrum access for certain duration, and the other is consumed for its own data transmission. By introducing the cooperation between PU and SU, the wireless-powered SU can achieve much higher throughput than other benchmark schemes while guaranteeing a fixed throughput requirement of the PU.

Instead of the time switching-based spectrum sharing scheme proposed in [5], the wireless-powered SU can also choose to operate in a block-wise EH manner. Reference [6] investigated such a scenario, where the SU implicitly overhears the primary link's transmission and harvest RF energy from the received signals over several consecutive transmission blocks, and opportunistically relay the PU data when it has accumulated a certain amount of energy. In each cooperative transmission block, the incremental DF relaying was implemented to efficiently use the harvested energy. Specifically, when the primary link suffers from an outage, the SU uses a fraction of its harvested energy to re-transmit the primary data along with the PU using Alamouti coding, and transmits its own data using the remaining fraction of its accumulated energy. Otherwise, the SU only transmits its own information in the second phase of cooperative transmission blocks. With the optimal power allocation between primary data relaying and secondary data transmission at the SU, the CSS strategy proposed in [6] can substantially improve the system throughput for both primary and secondary data transmissions. However, it should be noted that the CSS strategy proposed in [6] has a higher operation complexity than the time switching-based one developed in [5]. This is because the wireless-powered SU only needs to perform the allocation of the harvested energy in each transmission block in the former protocol, while the harvested energy in the latter protocol is needed to be scheduled across multiple transmission blocks due to the energy accumulation process.

The performance gain introduced by the above RF energy transfer-enabled CSS can be compromised by the low efficiency of RF energy transfer, especially when the PU and wireless-powered SU are not so close to each other such that the amount of energy harvested by the SU from the PU's signal could be small due to severe propagation loss. To address this issue, in a very recent work [7], Xing et al. proposed to deploy dedicated energy access points, which can not only transfer energy wirelessly to the SU but also help forward the PU's signal together with the SU.

B. *WPCRN with Wireless-powered PU*

In an alternative configuration of WPCRN, the PU, instead of the SU, is solely powered by the RF energy. In this situation, the SU with abundant power supply is enabled to perform both energy and information cooperation with the PU. The resource complementarity exists between primary and secondary systems: the PU possesses the licensed spectrum, but lacks of energy, while the SU has adequate energy but has no spectrum. The SU can use its energy to help the PU's EH and information transmission to exchange for the spectrum. To be more specific, the SU can wirelessly transfer energy to the PU together with the PU's serving base station (BS) to enhance the EH efficiency at the PU in DL. Besides, it can help forward the PU's information to the BS to enhance the transmission robustness in the UL. As such, more fraction of licensed spectrum can be released for the secondary data transmission. Zhai *et al.* first studied a typical WPCRN with a pair of PUs and a pair of SUs, in which the PU is exclusively powered by RF energy [8]. The CSS was facilitated via the bandwidth allocation, while the energy cooperation was realized via the time allocation. The information cooperation was performed in the subsequent transmission block only when the primary data's direct transmission between PU and BS is unsuccessful. The block diagram of this CSS protocol is shown in Fig. 5 for two distinct cases. The simulation results in [8] demonstrated that the proposed CSS protocol can dramatically boost the throughput of primary network.

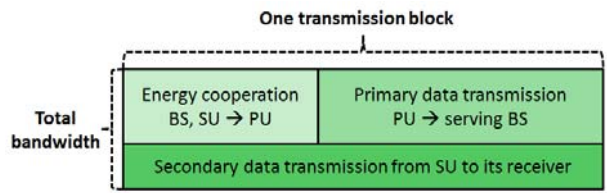

(a) Bandwith and time allocations when the primary data is correcly decoded at BS in the previous block

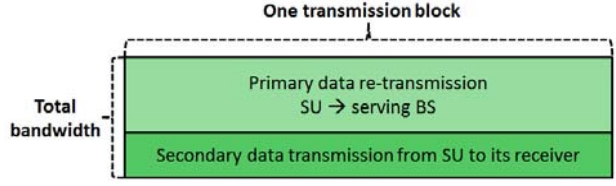

(b) Bandwith and time allocations when the primary data is erroneously decoded at BS in the previous block

Fig. 5 Block diagram of the EH-based CSS protocol for WPCRNs with wireless-powered PU proposed in [8].

Before ending this section, it is worth pointing out that in addition to the overlay spectrum sharing strategies introduced above, underlay cognitive relay network is another effective way to make use of the primary user's spectrum. Compared to the pure relay networks studied in Sec. II, each secondary node, no matter source or relay node, in underlay cognitive relay network needs to carefully control its transmit power to ensure that the interference caused by its transmission to the primary system does not exceed a predefined interference constraint. Due to space limitation, we refer interested readers to a very recent work [9] along this research direction and references therein.

## IV. COOPERATIVE JAMMING FOR WIRELESS-POWERED SECURE COMMUNICATIONS

The quick proliferations of wireless technologies have drawn unprecedented concerns on the security of wireless networks. Besides the cryptography approaches widely used at higher layers, the physical layer techniques, which leverage the inherent properties of wireless propagation channels (e.g., fading and interference) to achieve securer wireless communications, have been gradually becoming popular. Artificial noise-based approach has been recognized as an effective way to strengthen physical layer security. In this approach, a transmitter equipped with multiple antennas transmits secret information to the intended receiver, and at the same time, emits artificial noise to confound the surrounding malicious eavesdroppers. However, this approach cannot work appropriately when the information transmitter is equipped with single antenna due to size or cost limit. Motivated by this issue and inspired by the conventional cooperative relaying schemes, the cooperative jamming (CJ) technique was developed to reproduce the function of multiple transmit antennas. More specifically, when the single-antenna transmitter broadcasts its confidential information, single or multiple helper nodes (commonly termed friendly jammers) will simultaneously emit artificial noises to confuse the eavesdroppers.

The WPC technique has warranted the usage of wireless-powered jammer(s) to realize CJ for physical layer security. In this article, we term a secure communication protected by wireless-powered jammer(s) as a wireless-powered secure communication (WPSC). By incorporating wireless-powered jammer(s), the artificial noises can be generated by using the harvested energy, which can make the deployment of dedicated jammers more flexible.

Furthermore, it can stimulate more participations of normal nodes in the CJ process as they can help each other but avoid depleting their own batteries.

The idea of WPSCs was first proposed in [10], in which a typical four-node secure communication scenario was considered: in the presence of a passive eavesdropper, a source transmits its secret information to the intended destination under the protection of a wireless-powered friendly jammer. A new CJ protocol was designed for the considered WPSC. Since the friendly jammer is an energy-constrained node, it needs to first harvest energy from the source before it can generate artificial noises to assist the source to achieve secure communications. As such, it would be risky for the source to transmit secret information when the available amount of harvested energy at the jammer is relatively low as it cannot perform effective CJ to protect the information from eavesdropping. Motivated by this, in the proposed CJ protocol of [10], the source transmits secret information to its destination only when the link between them can support the required transmission rate and the jammer has accumulated a certain amount of energy to ensure an effective CJ; otherwise, the source will transfer RF energy to the jammer to further charge it for the usage in upcoming transmission blocks. The achievable throughput of the proposed CJ protocol was analyzed by characterizing its long-term behavior and deriving a closed-form expression of the secret information transmission probability. Furthermore, the rate parameters were optimized to attain the maximum throughput while ensuring a certain secrecy outage probability.

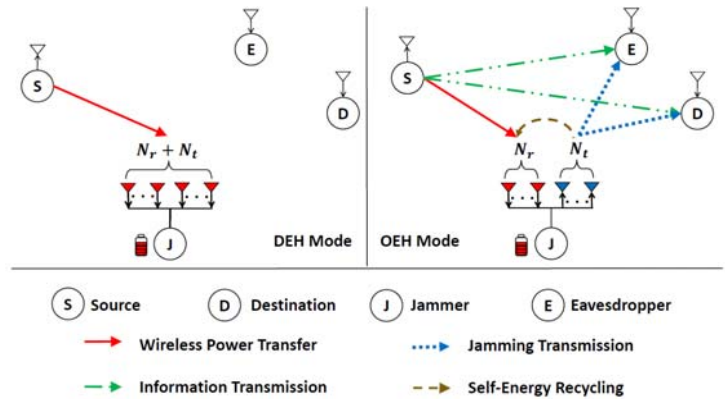

Fig. 6 Illustration of the AnJ protocol and its associated DEH and OEH modes [11]

Bi and Chen extended the concept of WPSC to another interesting scenario in [11], where a wireless-powered full-duplex (FD) jammer is employed to confuse passive eavesdropping. A novel CJ strategy, termed accumulate-and-jam (AnJ), was developed for the considered WPSC. In the proposed AnJ protocol, the system switches between two working modes (i.e., dedicated EH (DEH) mode and opportunistic EH (OEH) mode shown in Fig. 6) depending on the quality of the main link between source and destination and the amount of accumulated energy at the jammer. Compared with the half-duplex (HD) jammer adopted in [10], the FD wireless-powered jammer has two main merits: On one hand, the jammer can always scavenge energy from the source's signals, even when it is transmitting jamming signals. On the other hand, besides its main objective of confounding the eavesdropper, the jamming signals can also be utilized as a potential energy source via the self-loop channel among the transmitting antennas and receiving antennas of the jammer. As such, the FD jammer are expected to store more accumulated energy and execute more effective CJ than its HD counterpart in a long run. This deduction has been validated by the simulation results given in [11]. Some practical issues, such as finite battery capacity and imperfect channel state information, were incorporated into the system performance evaluation framework. Note that similar benefits of FD can also be harvested in WPCCNs and WPCRNs.

Reference [12] studied a secure AF relay communication scenario with multiple wireless-powered friendly jammers equipped with multiple antennas. To realize the information relaying, each transmission block is divided into two phases. In the first phase, the source sends information-bearing signals to the relay, which are treated as RF energy at the jammers and are deemed to be secure as no direct link between source and eavesdropper is assumed. In the second phase, the relay amplifies and forwards the source's information to the destination under the risk of being eavesdropped. To protect the information of the second hop, the wireless-powered jammers will consume the energy harvested during the first slot to perform CJ. The artificial noise covariance matrix for cooperative jamming and the AF beamforming matrix of the relay were jointly optimized to maximize the secrecy rate subject to the transmit power constraints of the AF relay and the wireless-powered jammers. Both perfect CSI and imperfect CSI scenarios were studied. It is worth pointing out that the framework developed in [12] failed to incorporate the inherent energy accumulation behavior of the jammer batteries as it presumes that the jammer batteries become empty at the beginning of each transmission block even though the harvested energy was not exhausted in the previous blocks.

In alternative applications, both the source and jammer can be wireless-powered nodes (e.g., they are both low-complexity IoT sensors). A very recent work [13] investigated such a system consisting of multiple wireless-powered users, one wireless-powered jammer, multiple eavesdroppers, and one hybrid access point (HAP). The HAP broadcasts energy-bearing signals to wirelessly charge the EH users and jammer; the users then use the harvested energy to transmit their data to the HAP one by one under the jammer's protection. The design aims of [13] were to maximize the minimum secrecy rate or minimize the maximum secrecy outage probability by jointly optimizing the time allocations among energy transfer and multiple users' transmission as well as the power allocation at the jammer.

V. FURTHER DISCUSSIONS AND POTENTIAL RESEARCH DIRECTIONS

In this section, we have some further discussions and identify some interesting research directions that can be further explored in the areas of cooperative WPC networks.

- **Helper selection and energy accumulation characterization in multi-helper WPC:** In most of the aforementioned cooperative strategies designed for WPC, only one helper node (i.e., relay/SU/jammer) was considered. In practice, there can be multiple helper nodes available to cooperate. How to select the helper node becomes a natural yet important question. To elaborate, we take a multi-relay WPCCN as an example, which consists of a source-destination pair as well as multiple wireless-powered relays. In conventional multi-relay cooperative networks, the relay selection can be conducted purely based on the dual-hop SNRs of the relays. For example, in the well-known max-min criterion, the relay with the maximum of minimum SNR of dual hops will be selected out to forward data such that the end-to-end SNR can be maximized. When it comes to the WPCCNs, since the relays are wireless-powered, their residual energy in the battery is a new key factor to be considered in relay selection. Furthermore, there is an inherent energy accumulation process when relay selection is performed in multi-relay WPCCN. Specifically, the non-selected relays can harvest energy from source signal and accumulate it for future usage. In this sense, the energy accumulation processes of multiple relays are tangled together and highly affected by the relay selection policy. There have been some initial efforts along this direction (e.g., [14] and references therein). However, what is the optimal helper selection criterion and how to properly characterize the energy accumulation process to analyze the system performance (e.g., diversity order) are still open problems.

- **Application-aware cooperative strategy design:** Most existing cooperative strategies for WPC in the literature were designed for general system setups. However, the specific applications of WPC substantially affect the characteristics and requirements for energy harvesting and cooperative strategies. For example, in wireless-powered cooperative sensor networks, sensors are equipped with limited energy and boosting the power supply to them is the most critical design objective. On the other hand, in some applications with user mobility (e.g., the energy access points can move), the cooperative strategies and their associated resource allocation should be re-examined in the sense that the energy consumed for wireless charging, information cooperation and movements as well as the moving trajectory and the stopping points should be jointly optimized. We can see from these two examples that more efforts and attentions should be paid on application-aware designs of cooperative WPC to pave the way for its implementations in real world.

- **Incentive mechanism design for cooperative WPC:** It was implicitly assumed in the literature that the helper nodes (e.g., relay/SU/jammer) are voluntary and willing to help once requested. This assumption is valid if all the involved nodes are operated by the same operator/service provider. However, the helper nodes can belong to a third party in practice. In this case, these helper nodes may expect some incentives (e.g., monetary rewards or free data offloading) to compensate the consumption of their own resources for cooperation. For example, a nearby mobile phone may be willing to help an IoT access point charge its associated IoT sensors to exchange for free data offloading. As such, effective incentive mechanisms should be properly designed to ensure that the cooperation can be performed efficiently among rational entities to significantly improve the network performance.

- **Ambient backscatter-based cooperative WPC:** Recently, ambient backscatter has emerged as a promising technology for low-power wireless communications, wherein the information is modulated and transmitted by modifying and reflecting ambient RF signals. The power consumption of a typical backscatter can be as low as 1 micro-Watt since the most power-hungry RF components (e.g., upconverter and power amplifier) are no longer needed in information transmission. In this sense, ambient backscatter can ideally integrate with WPC because the harvested energy at wireless-powered devices are normally quite limited. However, the communication range of ambient backscatter has been demonstrated to be too small for large-scale implementation. Designing new cooperative strategies for ambient backscatter-based WPC to enlarge its effective communication range is an interesting and important research direction.

- **Efficient cooperative schemes for ultra-reliable short-packet WPC:** The vast majority of existing work on WPC applied the Shannon capacity bound in their system design and performance analysis. An implicit assumption behind the Shannon formula is that the system is allowed to use an extremely long (or even infinite) packet length such that an arbitrarily low error rate can be realized at the receiver side. However, for most applications of WPC, the data is normally delivered in short packets with finite blocklength codes [15], where the packet error cannot be reduced to arbitrarily low. In this situation, user cooperation can be introduced to improve the system reliability by enhancing the signal-to-noise of each single hop. However, each hop needs to support a higher coding rate to ensure that the data can be conveyed to the destination using the same time duration as the non-cooperative system, which can deteriorate the reliability of each hop. This fundamental tradeoff makes whether to cooperate or not in short-packet WPC a non-trivial question to answer. More efforts are deserved to design efficient cooperative schemes to support ultra-reliable short-packet WPC.

## VI. CONCLUSIONS

In this article, we have overviewed the up-to-date progresses on the design and analysis of cooperative strategies for wireless-powered communications (WPCs), which can promisingly boost the network performance in a wide range of network setups. More specifically, we have summarized the recently developed relaying protocols for wireless-powered cooperative communications, cooperative spectrum sharing schemes for wireless-powered cognitive radio networks, and cooperative jamming strategies towards wireless-powered secure communications. At last, several future research directions for cooperative strategies in WPC were identified and discussed. In a nutshell, this article is expected to provide an accessible and general overview on how to design the cooperative strategies towards efficient WPCs.